\documentclass[a4paper,11pt]{article}
\usepackage{pos}
\usepackage{titlesec}
\titlespacing*{\section}{0pt}{6pt}{5pt}
\titlespacing*{\subsection}{0pt}{6pt}{5pt}

\title{Overview of Cherenkov Telescope on-board EUSO-SPB2 for the Detection of Very-High-Energy Neutrinos}
 \ShortTitle{EUSO-SPB2 Cherenkov Telescope}

\author*[a]{Mahdi Bagheri}
\author[b]{Peter Bertone}
\author[c]{Ivan Fontane}
\author[a]{Eliza Gazda}
\author[d]{Eleanor G. Judd}
\author[e]{John F. Krizmanic}
\author[c]{Evgeny N. Kuznetsov}
\author[f]{Michael J. Miller}
\author[f]{Jane Nachtman}
\author[f]{Yasar Onel}
\author[a]{A. Nepomuk Otte}
\author[c,g]{Patrick J. Reardon}
\author[a]{Oscar Romero Matamala}
\author[a]{Andrew Wang}
\author[h]{Lawrence Wiencke}

\affiliation[a]{School of Physics \& Center for Relativistic Astrophysics, Georgia Institute of Technology,\\
		     837 State Street NW, Atlanta, GA 30332-0430, USA}
\affiliation[b]{Marshall Space Flight Center, Huntsville, AL, USA}
\affiliation[c]{The University of Alabama in Huntsville, Huntsville, AL, USA}
\affiliation[d]{UC Berkley, Space Sciences Laboratory, Berkeley, CA, USA}
\affiliation[e]{University of Maryland, Baltimore County, Baltimore, MD, USA}
\affiliation[f]{University of Iowa, Iowa City, IA, USA}
\affiliation[g]{Center for Applied Optics, Huntsville, AL, USA}
\affiliation[h]{Colorado School of Mines, Golden, CO, USA}

\forColl{JEM-EUSO}

\emailAdd{mbagheri31@gatech.edu}

\abstract{We present the status of the development of a Cherenkov telescope to be flown on a long-duration balloon flight, the Extreme Universe Space Observatory Super Pressure Balloon 2 (EUSO-SPB2). EUSO-SPB2 is an approved NASA balloon mission that is planned to fly in 2023 and is a precursor of the Probe of Extreme Multi-Messenger Astrophysics (POEMMA), a candidate for an Astrophysics probe-class mission. The purpose of the Cherenkov telescope on-board EUSO-SPB2 is to classify known and unknown sources of backgrounds for future space-based neutrino detectors. Furthermore, we will use the Earth-skimming technique to search for Very-High-Energy (VHE) tau neutrinos below the limb (E > 10 PeV) and observe air showers from cosmic rays above the limb. The 0.785 $m^2$ Cherenkov telescope is equipped with a 512-pixel SiPM camera covering a 12.8° x 6.4° (Horizontal $\times$ Vertical) field of view. The camera signals are digitized with a 100 MS/s readout system. In this paper, we discuss the status of the telescope development, the camera integration, and simulation studies of the camera response.}

\FullConference{37$^{\rm{th}}$ International Cosmic Ray Conference (ICRC 2021)\\
		July 12th -- 23rd, 2021\\
		Online -- Berlin, Germany}

\begin{document}
\maketitle

\section{Introduction}
\begin{figure}[t]
	\centering
	\includegraphics[width=0.9\textwidth]{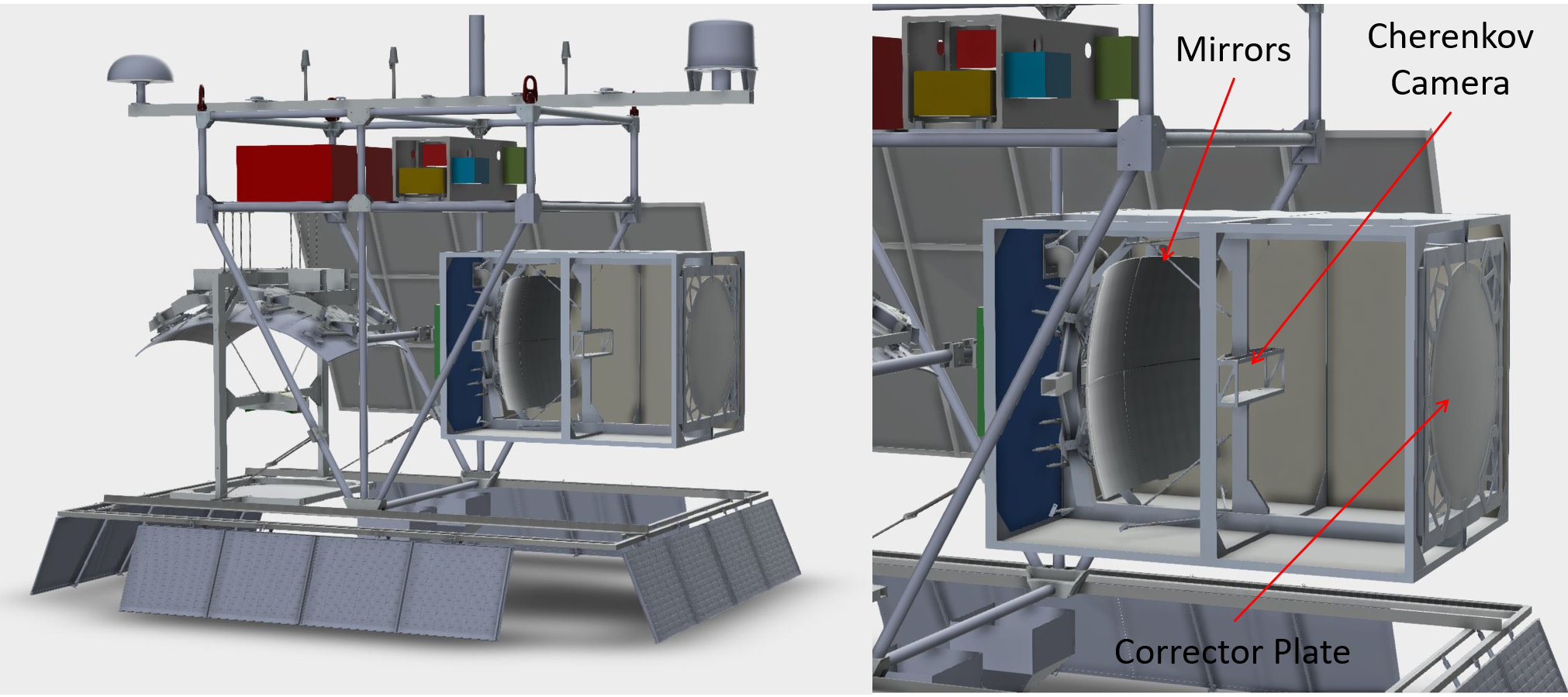}
	\caption{Left: EUSO-SPB2 payload, with the Fluorescence telescope pointing down for detecting UHECR and the Cherenkov telescope pointing at the Earth limb to detect VHE-neutrinos below-the-limb and UHECR via Cherenkov radiation above-the-limb. Right: Close-up of the Cherenkov telescope with the corrector plate.}
	\label{fig:Telescope}
\end{figure}

Multi-Messenger Astronomy is a rapidly growing field in part due to the detection of an astrophysical neutrino flux with IceCube \cite{IceCube2018} and the observation of gravitational waves with LIGO \cite{LIGO2016}. We are working towards opening another Multi-Messenger window to the universe, the Very-High-Energy (VHE) neutrino window (>10 PeV) with the Probe of Extreme Multi-Messenger Astrophysics (POEMMA) \cite{POEMMA-JCAP}. POEMMA is designed to study the flux and composition of Ultra-High Energy Cosmic Rays (UHECR) above 10$^{10.8}$ GeV and the flux of VHE neutrinos above 10$^7$ GeV. The precursor to POEMMA is the Extreme Universe Space Observatory Super Pressure Balloon 2 (EUSO-SPB2), which carries a Fluorescence Telescope and a Cherenkov Telescope. A rendering of the EUSO-SPB2 payload and the Cherenkov telescope are shown in Figure \ref{fig:Telescope}, left and right, respectively. In this paper, we report on the status of the on-board Cherenkov telescope.

EUSO-SPB2 will fly at an altitude of about 33 km. From that altitude, the Cherenkov telescope will, for the first time, observe a volume of 10° above and below the limb. With these observations, we will study the ambient background photon fields, also known as Night-Sky-Background (NSB), integrated over the Cherenkov telescope's spectral response and at different timescales. NSB refers to the residual light that is present in the night sky during dark, moonless nights, and is mostly caused by air-glow, direct and scattered starlight and zodiacal light \cite{BENN1998503}. These measurements will provide critical input for realistic predictions of VHE-neutrino observations with POEMMA. 

At times when the telescope is pointed at or below the limb, we will also search for Earth-skimming VHE-neutrinos. Earth-skimming VHE tau-neutrinos, i.e., VHE neutrinos entering the Earth under a shallow angle, have a high probability of interacting and producing a tau-lepton that emerges from the ground. The emerging tau then decays in the atmosphere starting an air shower with billions of charged particles. A significant amount of the particles' energy is converted into optical Cherenkov emission that is radiated into a narrow cone centered around the shower axis \cite{AlvarezMuniz2019,Cummings2021}. A Cherenkov telescope that happens to be inside the cone collects a fraction of the light and projects it onto the telescope's camera where the air shower is imaged. The offline analysis reconstructs the arrival direction and the energy of the neutrino based on the recorded image.

When the telescope is pointed above the limb, we anticipate imaging Extensive Air Showers (EAS) initiated by cosmic rays in the upper atmosphere. We expect that primary cosmic rays with energies above 1 PeV could result in as many as 100 detected events per hour \cite{aboveLimb}. These measurement will be helpful in evaluating the performance of the telescope. The Cherenkov telescope will also be capable of rotating towards transient astrophysical sources to search for neutrinos coming from them, although the probability of detecting such events is admittedly very low \cite{ToO_poemma}.

\section{Telescope Optics}

\begin{figure}[b]
	\centering
	\includegraphics[width=0.9\textwidth]{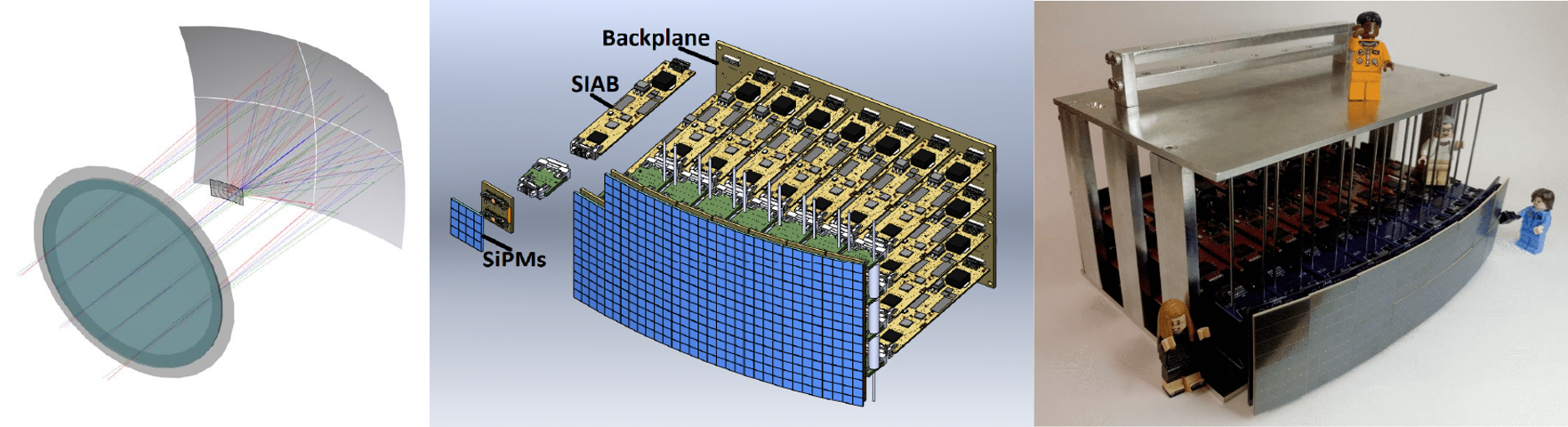}
	\caption{Left: Schmidt catadioptric optics of the Cherenkov telescope. The mirrors focus the light onto the curved focal plane located between the corrector plate and the mirror. Middle: CAD drawing of the camera without housing. Right: Half assembled camera (Lego figures for scale).}
	\label{fig:Mirrors}
\end{figure}

The Cherenkov telescope optics is based on a Schmidt catadioptric system with a focal length of 860 mm (Figure \ref{fig:Mirrors}, left). Aberrations are controlled with a corrector lens at the entrance of the telescope. The 1 m$^2$ light collection area is segmented into four identical mirrors. Taking into account the shadowing from the camera, transmission losses from the corrector plate, and reflection/scattering losses at the mirror, the resulting effective aperture area is 0.785 m$^2$. The camera focal plane is curved with a radius of 850 mm. In this setup, 90$\%$ of the light from a point source at infinity is contained in a 3 mm diameter circle. A unique feature of the telescope is its bi-focal optics, which projects the image twice on the camera with a horizontal offset of 12 mm. The split is achieved by rotating the optical axis of the lower and upper row of mirrors 0.4$^{\circ}$ relative to each other \cite{Mirrors}.

\section{Camera}
\subsection{Focal plane}

The focal plane of the Cherenkov telescope is instrumented with a 512 pixel silicon photomultiplier (SiPM) camera yielding an overall field of view of 12.8° $\times$ 6.4° in the horizontal and vertical, respectively. The size of a pixel is 6.4 mm $\times$ 6.4 mm and 16 pixels are grouped into a 4 $\times$ 4 matrix. The SiPM matrices are of the type S14521-6050AN-04 from Hamamatsu. The camera is populated with a total of 32 such matrices. The chosen SiPM has a broad sensitivity from 200 nm to 1000 nm with a peak photon detection efficiency of 50$\%$ at 450 nm. At the operating voltage, direct optical cross-talk is only 1.5$\%$ and the temperature dependence of the gain is only $\sim$ 0.5$\%$/°C.

The SiPM is an electrically and mechanically rugged and highly-sensitive single photon resolving photon detector. It is used in a wide range of astroparticle experiments. The wide spectral response of SiPMs reaches into the IR which is ideal for our purpose. Because of absorption and scattering, only the red components of the Cherenkov light arrive at the telescope from far-away showers, like the ones expected from Earth-skimming neutrinos.

\subsection{Front-end electronics}

Figure \ref{fig:Mirrors}, middle shows a CAD drawing of the camera. The raw SiPM signal are routed into the Sensor Interface and Amplifier Board (SIAB) via two adapter boards. One SIAB serves one 4 $\times$ 4 matrix of SiPMs. A total of 32 SIABs interface with a backplane from where the SiPM signals are routed to the digitizer. The SIABs receive power and communication through the backplane. On the SIAB, two Multipurpose Integrated Circuit (MUSIC) chips shape and amplify the SiPM signals. The MUSIC chip is an 8-channel, low-power Application Specific Integrated Circuit (ASIC) designed explicitly for SiPM applications in Cherenkov telescopes \cite{Gomez}. It provides a 9-bit Digital to Analog Converter (DAC) to adjust the SiPM bias voltage and a current-monitor output for each SiPM channel. We digitize the current monitor output on the SIAB with a 24-bit Analog to Digital Converter (ADC). A microcontroller (ATMEGA328P), also on the SIAB, communicates with the MUSIC chips and the ADC via Serial Peripheral Interface (SPI). The microcontroller monitors the common bias of the SiPMs, the current flowing through each SiPM channel, and the temperature of the SiPMs with a thermistor mounted to the back of each SiPM matrix. The temperature measurements will be used to offline correct temperature-dependent gain drifts of the SiPMs. Figure \ref{fig:Mirrors}, right shows the camera with 16 SiPM matrices and SIABs installed.

\begin{figure}[b]
	\centering
	\includegraphics[width=\textwidth]{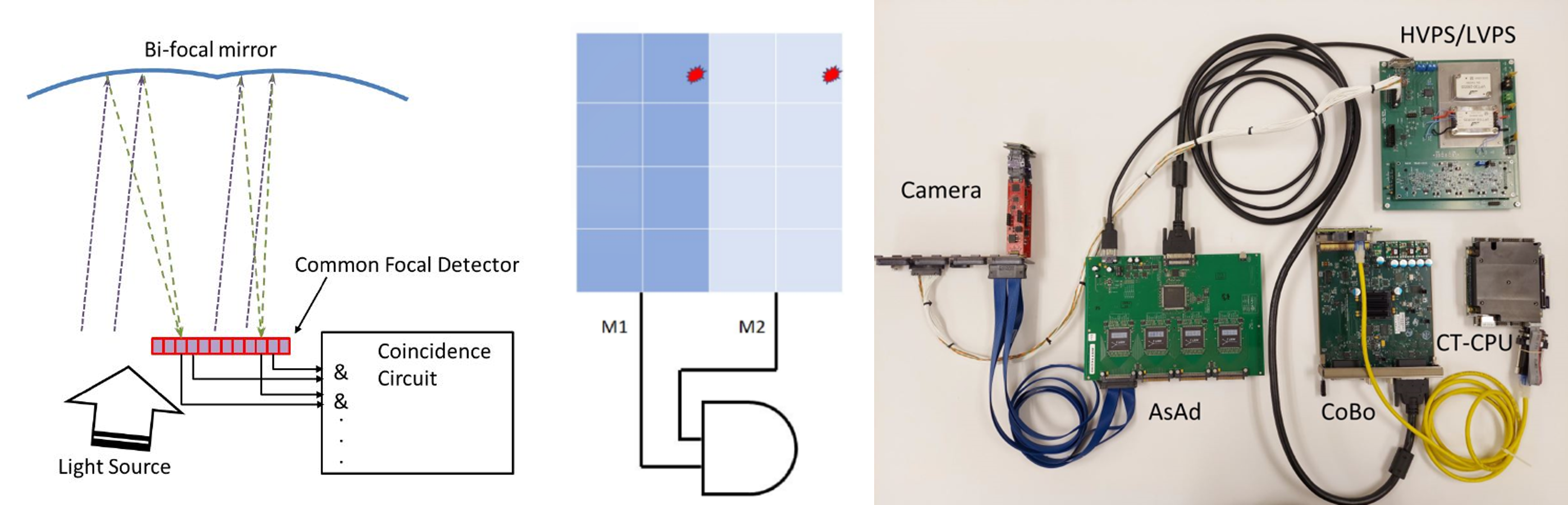}
	\caption{Left and Middle: The bi-focal optics duplicates the image and projects it to two separate locations in the camera. The telescope readout is triggered if two pixels separated by the bi-focal split record a signal within 100 ns. Right: Partially assembled camera and the readout electronics.}
	\label{fig:Trigger_Logic}
\end{figure}

\subsection{Trigger Logic}

The readout of the SiPM signals is initiated whenever the bi-focal trigger condition is met. Typical air-showers induced by an Earth-skimming tau-neutrino would usually produce a Cherenkov signal that illuminates only one pixel in the camera. With the bi-focal optics, the Cherenkov signal will be imaged into two pixels separated by one pixel. The splitting offset ensures that always SiPMs connected to two different MUSIC chips record most of the signal (See the red spots in M1 and M2 in the Figure \ref{fig:Trigger_Logic}, middle). The bi-focal trigger condition requires that two adjacent MUSIC chips must record a signal above the threshold of any of their internal leading-edge discriminators. The MUSIC discriminator outputs are routed to the trigger board where the coincidence (logic AND) with a 100 ns coincidence window is formed, as illustrated in Figure \ref{fig:Trigger_Logic}, middle. The bi-focal optics and the coincidence trigger circuit lower the energy threshold by effectively rejecting false triggers due to fluctuations in the NSB. The Trigger Board can also force trigger the readout at times derived from a 1 Hz GPS signal simultaneously with, internal triggers, external triggers and a flasher illuminating the camera. The flasher signals are used to monitor the performance of the telescope.

\subsection{Signal Digitization}

The shaped SiPM signals are routed through the backplane and micro-coaxial cables into the ASIC Support $\&$ Analog-Digital conversion (AsAd) boards. Each AsAd board has four ASIC for General Electronics for TPC (AGET) chips \cite{POLLACCO201881}. Implemented in the AGET are 64 switched capacitor arrays (SCA) that serve as analog ring buffers for the SiPM signals. The SCA is sampled with a rate of 100 MS/s providing a buffer depth of 5.12 $\mu$s. A favorable characteristic of the AGET is a low power consumption of < 10 mW per channel. When a readout command is received from the trigger, the signals stored in the SCAs are digitized with 12-bit resolution. We use two AsAd boards to digitize all 512 camera channels. The digitized signals are managed by a Concentration Board (CoBo). The CoBo is responsible for applying a time stamp, zero suppression and compression algorithms to the digitized signal. It also serves as a communication intermediary between the AsAd and the camera server, including the slow control signals and commands to the AsAd. Figure \ref{fig:Trigger_Logic}, right shows the camera, readout chain including the digitizer boards, CoBo and CPU. The total power consumption of the Cherenkov telescope is estimated to be about 180 watts during operation.

\subsection{Camera-Server Software}

For the camera server, we use a dual-core single board computer from RTD Embedded Technologies (CMA24CRD1700HR). The camera server receives the digitized SiPM signals and performs several management and monitoring tasks. It communicates with the Trigger board and the AGET digitizer through the Ethernet interface, with the front-end camera electronics (SIAB) through the System Management Bus (SMBus), with the housekeeping module through the serial interface, and with the Power Distribution Unit through CAN bus. The control software, developed in C++, relays the commands from the gondola computer to the corresponding processes running on the camera server using POSIX message queues. Replies are send back in a separate message queue. 

The acquisition of an event starts when the Trigger Board enables the trigger. The CoBo and the Trigger Board then wait for an event to trigger and store the digitized SiPM signals and the trigger information, respectively. After 5 minutes, the Trigger Board disables the trigger, and the server retrieves the raw data from the CoBo and Trigger Board and saves them on disk. The event builder running on the camera server merges the digitizer and trigger data into one data stream. A parallel process generates housekeeping files with information including the SiPM temperatures, bias voltages, currents, and other health and status data of the camera. Due to the limited bandwidth available for downloading the data during the flight, the event data will be prioritized. The housekeeping, health data, and air-shower candidate events are given the highest priority. Events close to the trigger threshold and data sampling background are given a lower priority.

\section{Camera Performance}
\subsection{Signal Chain Linearity}

\begin{figure}[t]
	\centering
	\includegraphics[width=0.8\textwidth]{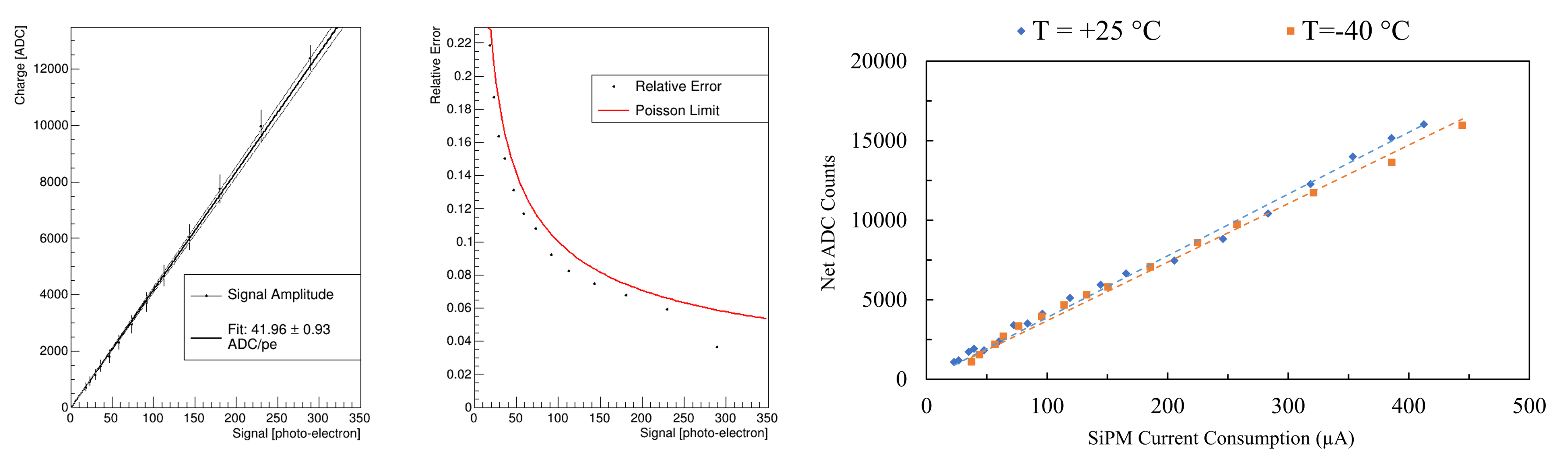}
	\caption{Left: Linearity and dynamic range of the full signal chain. Middle: Relative error compared with Poisson limit. Right: Response of the current monitoring system.}
	\label{fig:SignalChain}
\end{figure}

We have characterized the linearity and dynamic range of the full signal chain. Figure \ref{fig:SignalChain}, left shows the AGET digitized integral charge versus the number of detected photons, photoelectrons (PEs). For this measurement, we flashed the SiPMs with a picosecond laser and varied its intensity. The signal chain is linear up to 300 photoelectrons meeting the required dynamic range. The middle graph shows the relative error compared with the Poisson limit. The overall performance of the signal chain is not limited by the noise characteristics of the signal chain. Poisson fluctuations in the primary photon signal, which scale with 1/$\sqrt(PE)$ dominate the uncertainty in the recorded data. Electronic noise becomes dominant below a detected signal of 20 photoelectrons.

\subsection{Current Monitoring}

The SiPM currents are monitored on the SIAB using the corresponding monitoring signals provided by the MUSIC. In order to prevent damaging the input stages of the MUSIC chip, the SIAB microcontroller automatically turns off a MUSIC channel if the corresponding SiPM current crosses a threshold of 400\,$\mu$A. For that purpose, an ADC on the SIAB digitizes the current of each of the 16 pixels at a rate of 100 Hz with a resolution of 7.6 \textmu A. Figure \ref{fig:SignalChain} right shows the measured current-monitoring response of one channel at room temperature and at -40 °C. The microcontroller attempts to turn the channel back on after waiting for one minute and following a safety protocol to prevent the MUSIC chip from being damaged. The current-monitoring system's secondary purpose is to measure the photon background's intensity and its spatial and temporal variations. 

\section{Simulated Camera Response}

\begin{figure}[t]
	\centering
	\includegraphics[width=0.7\textwidth]{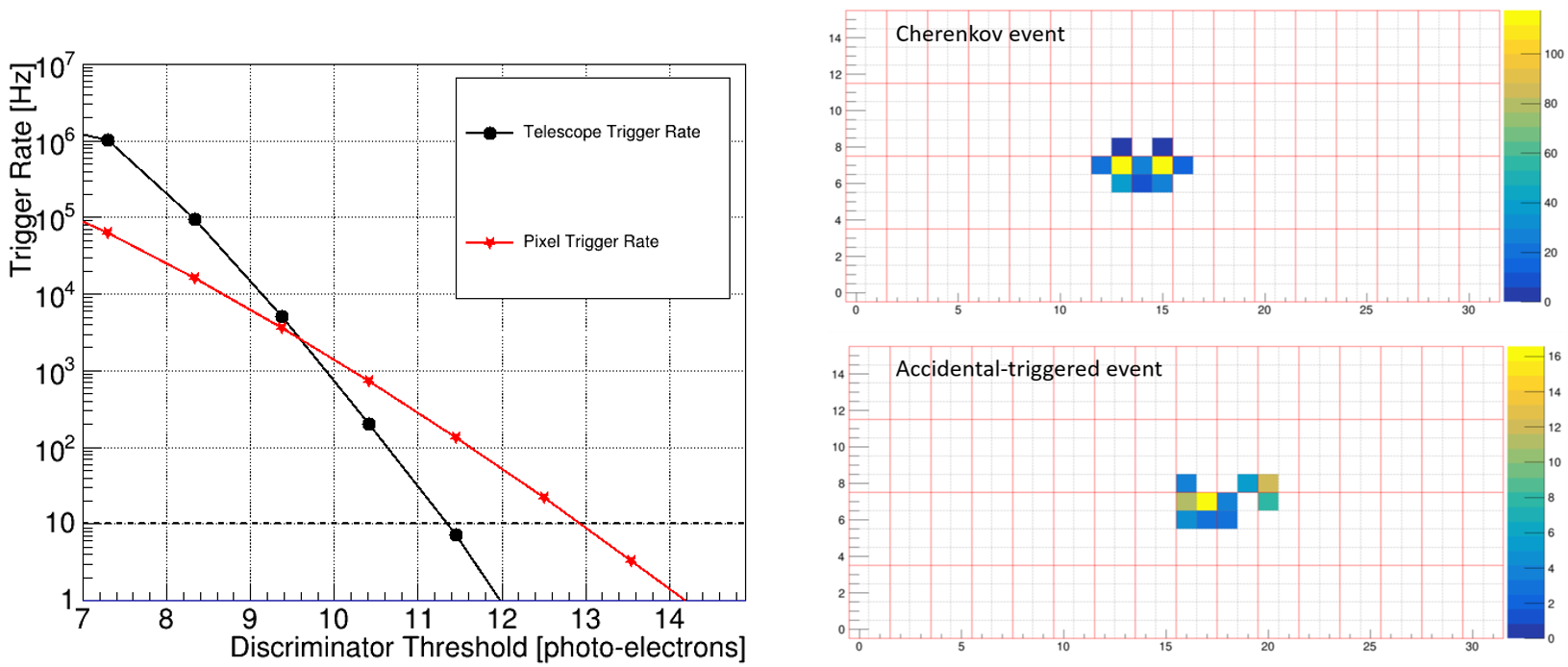}
	\caption{Left: Simulated trigger rates caused by background fluctuations for a single pixel in the camera and the 512-pixel camera after applying the bi-focal trigger. Right: Camera event display of a simulated air-shower event (top) and accidental-triggered event (bottom).}
    \label{fig:TrigRate}
\end{figure}

We studied the bi-focal trigger's performance, the reconstruction of the Cherenkov events, and the rejection of accidental events with a detailed model of the camera. The simulation model includes the proper response of the SiPMs, the camera electronics, the trigger, and the digitizer. The trigger threshold of the Cherenkov telescope is limited by fluctuations of the NSB, which results in accidental triggers. If the threshold is set too low, the trigger rate due to accidentals is too high and cannot be processed by the readout anymore. That situation is avoided by raising the trigger threshold of the readout. To estimate the camera's trigger threshold, we simulated the trigger assuming an expected NSB intensity of 3.7 $\times$ 10$^6$ pe/s/mm$^2$/sr \cite{BENN1998503} and setting different discriminator thresholds in the simulation. The trigger scan curves are  shown in Figure \ref{fig:TrigRate}, left for a single pixel and the whole camera. With a discriminator threshold of about 11 photoelectrons, the single pixel trigger is several hundred triggers per second, at the same time, the camera trigger rate after enforcing the coincidence trigger drops below the required 10 Hz to not saturate the data acquisition system.

\begin{figure}[t]
	\centering
	\includegraphics[width=0.8\textwidth]{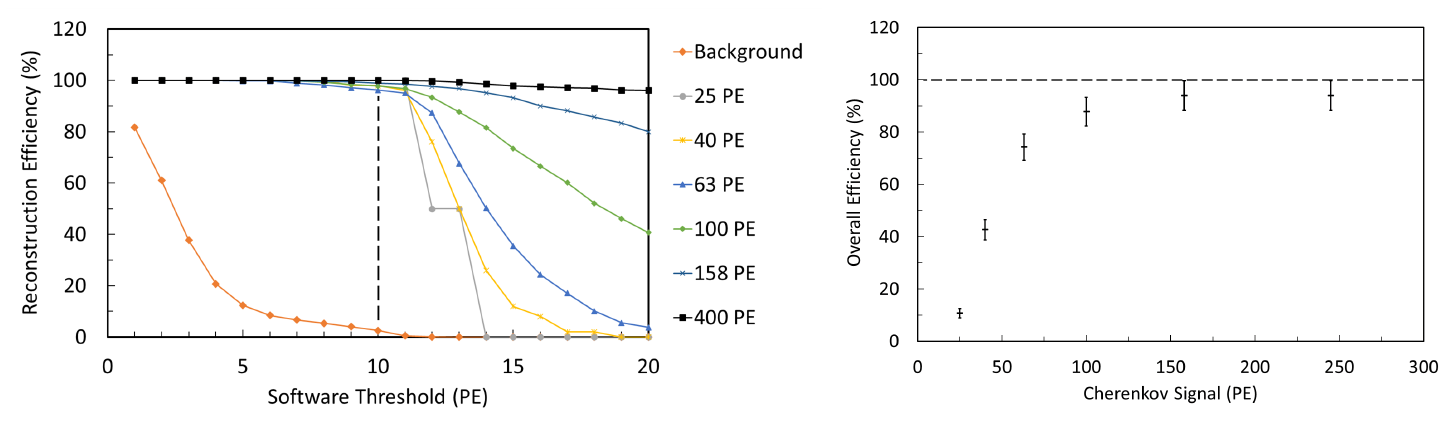}
	\caption{Left: Reconstruction efficiency of Cherenkov events compared to accidentals events (orange) for different photoelectron (PE) intensities. Right: Combined trigger and reconstruction efficiency of Cherenkov events as a function of the detected Cherenkov signal.}
	\label{fig:RecEff}
\end{figure}

The trigger logic allows any two pixels with random fluctuations to trigger the readout as long as the two pixels are connected to different but adjacent MUSIC chips. An example of such a noise event is shown in the bottom right panel of Figure \ref{fig:TrigRate}. An example of a simulated air-shower event is shown in the right top panel. In the case of a Cherenkov event, the bi-focal optics splits the Cherenkov light in the horizontal direction between two pixels separated by one pixel due to the bi-focal optics. In an accidental event no such topology is evident because of the random distribution of the NSB fluctuations. We make use of this topological difference in the event reconstruction by requiring an event to have a minimum signal in a pair of pixels separated by one pixel in the horizontal. The reconstruction algorithm scans all pairs of pixels that could have triggered the readout and if it finds a pair which meets the above condition, the event is accepted. Setting a 10-photoelectrons threshold in the analysis, the algorithm rejects 90$\%$ of accidentals and retains 95$\%$ of the Cherenkov events if the combined Cherenkov signal in all pixels is more than 25 photo-electrons as shown in Figure \ref{fig:RecEff}, left. The overall reconstruction efficiency, shown in the right panel of Figure \ref{fig:RecEff}, demonstrates that a Cherenkov event with more than 50 photoelectrons will be triggered and reconstructed with an efficiency of 50$\%$ or more. The efficiency is dominated by the trigger efficiency and Poisson fluctuations of the Cherenkov signal. Air-shower simulations show that a typical VHE-neutrino event in the upper atmosphere will generate a detectable signal of more than 50 photoelectrons in the EUSO-SPB2 telescope \cite{Cummings2021}. We are thus confident that the performance of the Cherenkov telescope is not limited by the camera.

\section{Conclusion}
We have developed the Cherenkov telescope that will fly as part of EUSO-SPB2 to study ambient photon field and background fluctuations and to attempt, for the first time, to detect air showers produced by VHE-tau neutrinos below-the-limb and cosmic rays above-the-limb. The SiPMs, camera electronics and readout system are tested and will undergo field tests later this year in preparation of the balloon flight, which is scheduled for 2023 from Wanaka, New Zealand.

\bigbreak

\bibliographystyle{JHEP}
\bibliography{references.bib}

\providecommand{\href}[2]{#2}\begingroup\raggedright\begin{thebibliography}{10}

\bibitem{IceCube2018}
{\scshape ICECUBE} collaboration, \emph{Neutrino emission from the direction of
  the blazar txs 0506+056 prior to the icecube-170922a alert},
  \href{https://doi.org/10.1126/science.aat2890}{\emph{Science} {\bfseries 361}
  (2018) 147}.

\bibitem{LIGO2016}
{\scshape LIGO Scientific Collaboration and Virgo} collaboration,
  \emph{Observation of gravitational waves from a binary black hole merger},
  \href{https://doi.org/10.1103/PhysRevLett.116.061102}{\emph{Phys. Rev. Lett.}
  {\bfseries 116} (2016) 061102}.

\bibitem{POEMMA-JCAP}
A.~Olinto et~al., \emph{The {POEMMA} ({Probe of Extreme Multi-Messenger
  Astrophysics}) observatory},
  \href{https://doi.org/10.1088/1475-7516/2021/06/007}{\emph{Journal of
  Cosmology and Astroparticle Physics} {\bfseries 2021} (2021) 007}.

\bibitem{BENN1998503}
C.~Benn and S.~Ellison, \emph{Brightness of the night sky over la palma},
  \href{https://doi.org/https://doi.org/10.1016/S1387-6473(98)00062-1}{\emph{New
  Astronomy Reviews} {\bfseries 42} (1998) 503}.

\bibitem{AlvarezMuniz2019}
J.~Alvarez-Mu\~niz et~al., \emph{Erratum: Comprehensive approach to tau-lepton
  production by high-energy tau neutrinos propagating through the earth [phys.
  rev. d 97, 023021 (2018)]},
  \href{https://doi.org/10.1103/PhysRevD.99.069902}{\emph{Phys. Rev. D}
  {\bfseries 99} (2019) 069902}.

\bibitem{Cummings2021}
A.L.~Cummings et~al., \emph{Modeling of the tau and muon neutrino-induced
  optical cherenkov signals from upward-moving extensive air showers},
  \href{https://doi.org/10.1103/PhysRevD.103.043017}{\emph{Phys. Rev. D}
  {\bfseries 103} (2021) 043017}.

\bibitem{aboveLimb}
A.~Cummings et~al., \emph{Modeling the optical cherenkov signals by cosmic ray
  extensive air showers directly observed from sub-orbital and orbital
  altitudes},  \href{https://arxiv.org/abs/arXiv:2105.03255v2}{{\ttfamily
  arXiv:2105.03255v2}}.

\bibitem{ToO_poemma}
T.M.~Venters et~al., \emph{Poemma's target-of-opportunity sensitivity to cosmic
  neutrino transient sources},
  \href{https://doi.org/10.1103/PhysRevD.102.123013}{\emph{Phys. Rev. D}
  {\bfseries 102} (2020) 123013}.

\bibitem{Mirrors}
V.~Kungel et~al., \emph{{EUSO-SPB2 Telescope Optics and Testing}},  in
  \emph{{Proceedings, 37th International Cosmic Ray Conference}},
  vol.~ICRC2021, p.~867, 2021.

\bibitem{Gomez}
S.~Gómez et~al., \emph{{MUSIC: An 8 channel readout ASIC for SiPM arrays}},
  in \emph{Optical Sensing and Detection IV}, vol.~9899, pp.~85 -- 94,
  International Society for Optics and Photonics, SPIE, 2016,
  \href{https://doi.org/10.1117/12.2231095}{https://doi.org/10.1117/12.2231095}.

\bibitem{POLLACCO201881}
E.~Pollacco et~al., \emph{Get: A generic electronics system for tpcs and
  nuclear physics instrumentation},
  \href{https://doi.org/https://doi.org/10.1016/j.nima.2018.01.020}{\emph{Nuclear
  Instruments and Methods in Physics Research Section A: Accelerators,
  Spectrometers, Detectors and Associated Equipment} {\bfseries 887} (2018)
  81}.

\end{thebibliography}\endgroup



\providecommand{\href}[2]{#2}\begingroup\raggedright\endgroup

\clearpage
\section*{Full Authors List: \Coll\ Collaboration}
\begin{sloppypar}
\scriptsize
\noindent
G.~Abdellaoui$^{ah}$, 
S.~Abe$^{fq}$, 
J.H.~Adams Jr.$^{pd}$, 
D.~Allard$^{cb}$, 
G.~Alonso$^{md}$, 
L.~Anchordoqui$^{pe}$,
A.~Anzalone$^{eh,ed}$, 
E.~Arnone$^{ek,el}$,
K.~Asano$^{fe}$,
R.~Attallah$^{ac}$, 
H.~Attoui$^{aa}$, 
M.~Ave~Pernas$^{mc}$,
%S.~Bacholle$^{pc}$, 
M.~Bagheri$^{ph}$,
J.~Bal\'az{$^{la}$, 
M.~Bakiri$^{aa}$, 
D.~Barghini$^{el,ek}$,
S.~Bartocci$^{ei,ej}$,
M.~Battisti$^{ek,el}$,
J.~Bayer$^{dd}$, 
B.~Beldjilali$^{ah}$, 
T.~Belenguer$^{mb}$,
N.~Belkhalfa$^{aa}$, 
R.~Bellotti$^{ea,eb}$, 
A.A.~Belov$^{kb}$, 
K.~Benmessai$^{aa}$, 
M.~Bertaina$^{ek,el}$,
P.F.~Bertone$^{pf}$,
P.L.~Biermann$^{db}$,
F.~Bisconti$^{el,ek}$, 
C.~Blaksley$^{ft}$, 
N.~Blanc$^{oa}$,
S.~Blin-Bondil$^{ca,cb}$, 
P.~Bobik$^{la}$, 
M.~Bogomilov$^{ba}$,
E.~Bozzo$^{ob}$,
S.~Briz$^{pb}$, 
A.~Bruno$^{eh,ed}$, 
K.S.~Caballero$^{hd}$,
F.~Cafagna$^{ea}$, 
G.~Cambi\'e$^{ei,ej}$,
D.~Campana$^{ef}$, 
J-N.~Capdevielle$^{cb}$, 
F.~Capel$^{de}$, 
A.~Caramete$^{ja}$, 
L.~Caramete$^{ja}$, 
P.~Carlson$^{na}$, 
R.~Caruso$^{ec,ed}$, 
M.~Casolino$^{ft,ei}$,
C.~Cassardo$^{ek,el}$, 
A.~Castellina$^{ek,em}$,
O.~Catalano$^{eh,ed}$, 
A.~Cellino$^{ek,em}$,
K.~\v{C}ern\'{y}$^{bb}$,  
M.~Chikawa$^{fc}$, 
G.~Chiritoi$^{ja}$, 
M.J.~Christl$^{pf}$, 
R.~Colalillo$^{ef,eg}$,
L.~Conti$^{en,ei}$, 
G.~Cotto$^{ek,el}$, 
H.J.~Crawford$^{pa}$, 
R.~Cremonini$^{el}$,
A.~Creusot$^{cb}$, 
A.~de Castro G\'onzalez$^{pb}$,  
C.~de la Taille$^{ca}$, 
L.~del Peral$^{mc}$, 
A.~Diaz Damian$^{cc}$,
R.~Diesing$^{pb}$,
P.~Dinaucourt$^{ca}$,
A.~Djakonow$^{ia}$, 
T.~Djemil$^{ac}$, 
A.~Ebersoldt$^{db}$,
T.~Ebisuzaki$^{ft}$,
L.~Eliasson$^{na}$, 
J.~Eser$^{pb}$,
F.~Fenu$^{ek,el}$, 
S.~Fern\'andez-Gonz\'alez$^{ma}$, 
S.~Ferrarese$^{ek,el}$,
G.~Filippatos$^{pc}$, 
 W.I.~Finch$^{pc}$
C.~Fornaro$^{en,ei}$,
M.~Fouka$^{ab}$, 
A.~Franceschi$^{ee}$, 
S.~Franchini$^{md}$, 
C.~Fuglesang$^{na}$, 
T.~Fujii$^{fg}$, 
M.~Fukushima$^{fe}$, 
P.~Galeotti$^{ek,el}$, 
E.~Garc\'ia-Ortega$^{ma}$, 
D.~Gardiol$^{ek,em}$,
G.K.~Garipov$^{kb}$, 
E.~Gasc\'on$^{ma}$, 
E.~Gazda$^{ph}$, 
J.~Genci$^{lb}$, 
A.~Golzio$^{ek,el}$,
C.~Gonz\'alez~Alvarado$^{mb}$, 
P.~Gorodetzky$^{ft}$, 
A.~Green$^{pc}$,  
F.~Guarino$^{ef,eg}$, 
C.~Gu\'epin$^{pl}$,
A.~Guzm\'an$^{dd}$, 
Y.~Hachisu$^{ft}$,
A.~Haungs$^{db}$,
J.~Hern\'andez Carretero$^{mc}$,
L.~Hulett$^{pc}$,  
D.~Ikeda$^{fe}$, 
N.~Inoue$^{fn}$, 
S.~Inoue$^{ft}$,
F.~Isgr\`o$^{ef,eg}$, 
Y.~Itow$^{fk}$, 
T.~Jammer$^{dc}$, 
S.~Jeong$^{gb}$, 
E.~Joven$^{me}$, 
E.G.~Judd$^{pa}$,
J.~Jochum$^{dc}$, 
F.~Kajino$^{ff}$, 
T.~Kajino$^{fi}$,
S.~Kalli$^{af}$, 
I.~Kaneko$^{ft}$, 
Y.~Karadzhov$^{ba}$, 
M.~Kasztelan$^{ia}$, 
K.~Katahira$^{ft}$, 
K.~Kawai$^{ft}$, 
Y.~Kawasaki$^{ft}$,  
A.~Kedadra$^{aa}$, 
H.~Khales$^{aa}$, 
B.A.~Khrenov$^{kb}$, 
 Jeong-Sook~Kim$^{ga}$, 
Soon-Wook~Kim$^{ga}$, 
M.~Kleifges$^{db}$,
P.A.~Klimov$^{kb}$,
D.~Kolev$^{ba}$, 
I.~Kreykenbohm$^{da}$, 
J.F.~Krizmanic$^{pf,pk}$, 
K.~Kr\'olik$^{ia}$,
V.~Kungel$^{pc}$,  
Y.~Kurihara$^{fs}$, 
A.~Kusenko$^{fr,pe}$, 
E.~Kuznetsov$^{pd}$, 
H.~Lahmar$^{aa}$, 
F.~Lakhdari$^{ag}$,
J.~Licandro$^{me}$, 
L.~L\'opez~Campano$^{ma}$, 
F.~L\'opez~Mart\'inez$^{pb}$, 
S.~Mackovjak$^{la}$, 
M.~Mahdi$^{aa}$, 
D.~Mand\'{a}t$^{bc}$,
M.~Manfrin$^{ek,el}$,
L.~Marcelli$^{ei}$, 
J.L.~Marcos$^{ma}$,
W.~Marsza{\l}$^{ia}$, 
Y.~Mart\'in$^{me}$, 
O.~Martinez$^{hc}$, 
K.~Mase$^{fa}$, 
R.~Matev$^{ba}$, 
J.N.~Matthews$^{pg}$, 
N.~Mebarki$^{ad}$, 
G.~Medina-Tanco$^{ha}$, 
A.~Menshikov$^{db}$,
A.~Merino$^{ma}$, 
M.~Mese$^{ef,eg}$, 
J.~Meseguer$^{md}$, 
S.S.~Meyer$^{pb}$,
J.~Mimouni$^{ad}$, 
H.~Miyamoto$^{ek,el}$, 
Y.~Mizumoto$^{fi}$,
A.~Monaco$^{ea,eb}$, 
J.A.~Morales de los R\'ios$^{mc}$,
M.~Mastafa$^{pd}$, 
S.~Nagataki$^{ft}$, 
S.~Naitamor$^{ab}$, 
T.~Napolitano$^{ee}$,
J.~M.~Nachtman$^{pi}$
A.~Neronov$^{ob,cb}$, 
K.~Nomoto$^{fr}$, 
T.~Nonaka$^{fe}$, 
T.~Ogawa$^{ft}$, 
S.~Ogio$^{fl}$, 
H.~Ohmori$^{ft}$, 
A.V.~Olinto$^{pb}$,
Y.~Onel$^{pi}$
G.~Osteria$^{ef}$,  
A.N.~Otte$^{ph}$,  
A.~Pagliaro$^{eh,ed}$, 
W.~Painter$^{db}$,
M.I.~Panasyuk$^{kb}$, 
B.~Panico$^{ef}$,  
E.~Parizot$^{cb}$, 
I.H.~Park$^{gb}$, 
B.~Pastircak$^{la}$, 
T.~Paul$^{pe}$,
M.~Pech$^{bb}$, 
I.~P\'erez-Grande$^{md}$, 
F.~Perfetto$^{ef}$,  
T.~Peter$^{oc}$,
P.~Picozza$^{ei,ej,ft}$, 
S.~Pindado$^{md}$, 
L.W.~Piotrowski$^{ib}$,
S.~Piraino$^{dd}$, 
Z.~Plebaniak$^{ek,el,ia}$, 
A.~Pollini$^{oa}$,
E.M.~Popescu$^{ja}$, 
R.~Prevete$^{ef,eg}$,
G.~Pr\'ev\^ot$^{cb}$,
H.~Prieto$^{mc}$, 
M.~Przybylak$^{ia}$, 
G.~Puehlhofer$^{dd}$, 
M.~Putis$^{la}$,   
P.~Reardon$^{pd}$, 
M.H..~Reno$^{pi}$, 
M.~Reyes$^{me}$,
M.~Ricci$^{ee}$, 
M.D.~Rodr\'iguez~Fr\'ias$^{mc}$, 
O.F.~Romero~Matamala$^{ph}$,  
F.~Ronga$^{ee}$, 
%I.~Rusinov$^{ba}$,
M.D.~Sabau$^{mb}$, 
G.~Sacc\'a$^{ec,ed}$, 
G.~S\'aez~Cano$^{mc}$, 
H.~Sagawa$^{fe}$, 
Z.~Sahnoune$^{ab}$, 
A.~Saito$^{fg}$, 
N.~Sakaki$^{ft}$, 
H.~Salazar$^{hc}$, 
J.C.~Sanchez~Balanzar$^{ha}$,
J.L.~S\'anchez$^{ma}$, 
A.~Santangelo$^{dd}$, 
A.~Sanz-Andr\'es$^{md}$, 
M.~Sanz~Palomino$^{mb}$, 
O.A.~Saprykin$^{kc}$,
F.~Sarazin$^{pc}$,
M.~Sato$^{fo}$, 
A.~Scagliola$^{ea,eb}$, 
T.~Schanz$^{dd}$, 
H.~Schieler$^{db}$,
P.~Schov\'{a}nek$^{bc}$,
V.~Scotti$^{ef,eg}$,
M.~Serra$^{me}$, 
S.A.~Sharakin$^{kb}$,
H.M.~Shimizu$^{fj}$, 
K.~Shinozaki$^{ia}$, 
T.~Shirahama$^{fn}$,
J.F.~Soriano$^{pe}$,
A.~Sotgiu$^{ei,ej}$,
I.~Stan$^{ja}$, 
I.~Strharsk\'y$^{la}$, 
N.~Sugiyama$^{fj}$, 
D.~Supanitsky$^{ha}$, 
M.~Suzuki$^{fm}$, 
J.~Szabelski$^{ia}$,
N.~Tajima$^{ft}$, 
T.~Tajima$^{ft}$,
Y.~Takahashi$^{fo}$, 
M.~Takeda$^{fe}$, 
Y.~Takizawa$^{ft}$, 
M.C.~Talai$^{ac}$, 
Y.~Tameda$^{fu}$, 
C.~Tenzer$^{dd}$,
S.B.~Thomas$^{pg}$, 
O.~Tibolla$^{he}$,
L.G.~Tkachev$^{ka}$,
T.~Tomida$^{fh}$, 
N.~Tone$^{ft}$, 
S.~Toscano$^{ob}$, 
M.~Tra\"{i}che$^{aa}$, 
%R.~Tsenov$^{ba}$, 
Y.~Tsunesada$^{fl}$, 
K.~Tsuno$^{ft}$,  
S.~Turriziani$^{ft}$, 
Y.~Uchihori$^{fb}$, 
O.~Vaduvescu$^{me}$, 
J.F.~Vald\'es-Galicia$^{ha}$, 
P.~Vallania$^{ek,em}$,
L.~Valore$^{ef,eg}$,
G.~Vankova-Kirilova$^{ba}$, 
T.~M.~Venters$^{pj}$,
C.~Vigorito$^{ek,el}$, 
L.~Villase\~{n}or$^{hb}$,
B.~Vlcek$^{mc}$, 
P.~von Ballmoos$^{cc}$,
M.~Vrabel$^{lb}$, 
S.~Wada$^{ft}$, 
J.~Watanabe$^{fi}$, 
J.~Watts~Jr.$^{pd}$, 
R.~Weigand Mu\~{n}oz$^{ma}$, 
A.~Weindl$^{db}$,
L.~Wiencke$^{pc}$, 
M.~Wille$^{da}$, 
J.~Wilms$^{da}$,
D.~Winn$^{pm}$
T.~Yamamoto$^{ff}$,
J.~Yang$^{gb}$,
H.~Yano$^{fm}$,
I.V.~Yashin$^{kb}$,
D.~Yonetoku$^{fd}$, 
S.~Yoshida$^{fa}$, 
R.~Young$^{pf}$,
I.S~Zgura$^{ja}$, 
M.Yu.~Zotov$^{kb}$,
A.~Zuccaro~Marchi$^{ft}$
}
\end{sloppypar}
\vspace*{.3cm}

%%\newpage
{
\footnotesize
\noindent
% Algeria (Dezember 2013) - 7 institutes
$^{aa}$ Centre for Development of Advanced Technologies (CDTA), Algiers, Algeria \\
$^{ab}$ Dep. Astronomy, Centre Res. Astronomy, Astrophysics and Geophysics (CRAAG), Algiers, Algeria \\
$^{ac}$ LPR at Dept. of Physics, Faculty of Sciences, University Badji Mokhtar, Annaba, Algeria \\
$^{ad}$ Lab. of Math. and Sub-Atomic Phys. (LPMPS), Univ. Constantine I, Constantine, Algeria \\
$^{af}$ Department of Physics, Faculty of Sciences, University of M'sila, M'sila, Algeria \\
$^{ag}$ Research Unit on Optics and Photonics, UROP-CDTA, S\'etif, Algeria \\
$^{ah}$ Telecom Lab., Faculty of Technology, University Abou Bekr Belkaid, Tlemcen, Algeria \\
% Bulgaria ready (02042012)  - 1 institutes 
$^{ba}$ St. Kliment Ohridski University of Sofia, Bulgaria\\
% Czech Republic (01072021) - 2 institutes
$^{bb}$ Joint Laboratory of Optics, Faculty of Science, Palack\'{y} University, Olomouc, Czech Republic\\
$^{bc}$ Institute of Physics of the Czech Academy of Sciences, Prague, Czech Republic\\
% France ready (02042012)  - 3 institutes 
$^{ca}$ Omega, Ecole Polytechnique, CNRS/IN2P3, Palaiseau, France\\
$^{cb}$ Universit\'e de Paris, CNRS, AstroParticule et Cosmologie, F-75013 Paris, France\\
$^{cc}$ IRAP, Universit\'e de Toulouse, CNRS, Toulouse, France\\
% Germany ready (01072021)  - 5 institutes
$^{da}$ ECAP, University of Erlangen-Nuremberg, Germany\\
$^{db}$ Karlsruhe Institute of Technology (KIT), Germany\\
$^{dc}$ Experimental Physics Institute, Kepler Center, University of T\"ubingen, Germany\\
$^{dd}$ Institute for Astronomy and Astrophysics, Kepler Center, University of T\"ubingen, Germany\\
$^{de}$ Technical University of Munich, Munich, Germany\\
% Italy ready (01042012)  - 14 institutes 
$^{ea}$ Istituto Nazionale di Fisica Nucleare - Sezione di Bari, Italy\\
$^{eb}$ Universita' degli Studi di Bari Aldo Moro and INFN - Sezione di Bari, Italy\\
$^{ec}$ Dipartimento di Fisica e Astronomia "Ettore Majorana", Universita di Catania, Italy\\
$^{ed}$ Istituto Nazionale di Fisica Nucleare - Sezione di Catania, Italy\\
$^{ee}$ Istituto Nazionale di Fisica Nucleare - Laboratori Nazionali di Frascati, Italy\\
$^{ef}$ Istituto Nazionale di Fisica Nucleare - Sezione di Napoli, Italy\\
$^{eg}$ Universita' di Napoli Federico II - Dipartimento di Fisica "Ettore Pancini", Italy\\
$^{eh}$ INAF - Istituto di Astrofisica Spaziale e Fisica Cosmica di Palermo, Italy\\
$^{ei}$ Istituto Nazionale di Fisica Nucleare - Sezione di Roma Tor Vergata, Italy\\
$^{ej}$ Universita' di Roma Tor Vergata - Dipartimento di Fisica, Roma, Italy\\
$^{ek}$ Istituto Nazionale di Fisica Nucleare - Sezione di Torino, Italy\\
$^{el}$ Dipartimento di Fisica, Universita' di Torino, Italy\\
$^{em}$ Osservatorio Astrofisico di Torino, Istituto Nazionale di Astrofisica, Italy\\
$^{en}$ Uninettuno University, Rome, Italy\\
% Japan ready (30032012)  - 20 institutes 
$^{fa}$ Chiba University, Chiba, Japan\\ 
$^{fb}$ National Institutes for Quantum and Radiological Science and Technology (QST), Chiba, Japan\\ 
$^{fc}$ Kindai University, Higashi-Osaka, Japan\\ 
$^{fd}$ Kanazawa University, Kanazawa, Japan\\ 
$^{fe}$ Institute for Cosmic Ray Research, University of Tokyo, Kashiwa, Japan\\ 
$^{ff}$ Konan University, Kobe, Japan\\ 
$^{fg}$ Kyoto University, Kyoto, Japan\\ 
$^{fh}$ Shinshu University, Nagano, Japan \\
$^{fi}$ National Astronomical Observatory, Mitaka, Japan\\ 
$^{fj}$ Nagoya University, Nagoya, Japan\\ 
$^{fk}$ Institute for Space-Earth Environmental Research, Nagoya University, Nagoya, Japan\\ 
$^{fl}$ Graduate School of Science, Osaka City University, Japan\\ 
$^{fm}$ Institute of Space and Astronautical Science/JAXA, Sagamihara, Japan\\ 
$^{fn}$ Saitama University, Saitama, Japan\\ 
$^{fo}$ Hokkaido University, Sapporo, Japan \\ 
$^{fp}$ Osaka Electro-Communication University, Neyagawa, Japan\\ 
$^{fq}$ Nihon University Chiyoda, Tokyo, Japan\\ 
$^{fr}$ University of Tokyo, Tokyo, Japan\\ 
$^{fs}$ High Energy Accelerator Research Organization (KEK), Tsukuba, Japan\\ 
$^{ft}$ RIKEN, Wako, Japan\\
% Korea (02042012)  - 2 institutes
$^{ga}$ Korea Astronomy and Space Science Institute (KASI), Daejeon, Republic of Korea\\
$^{gb}$ Sungkyunkwan University, Seoul, Republic of Korea\\
% Mexico (02042012)  - 5 institutes
$^{ha}$ Universidad Nacional Aut\'onoma de M\'exico (UNAM), Mexico\\
$^{hb}$ Universidad Michoacana de San Nicolas de Hidalgo (UMSNH), Morelia, Mexico\\
$^{hc}$ Benem\'{e}rita Universidad Aut\'{o}noma de Puebla (BUAP), Mexico\\
$^{hd}$ Universidad Aut\'{o}noma de Chiapas (UNACH), Chiapas, Mexico \\
$^{he}$ Centro Mesoamericano de F\'{i}sica Te\'{o}rica (MCTP), Mexico \\
% Poland ready (01072021)  - 2 institutes
$^{ia}$ National Centre for Nuclear Research, Lodz, Poland\\
$^{ib}$ Faculty of Physics, University of Warsaw, Poland\\
% Romania ready (Jan 2015) - 1 institute 
$^{ja}$ Institute of Space Science ISS, Magurele, Romania\\
% Russia ready (30032012)  - 3 institutes 
$^{ka}$ Joint Institute for Nuclear Research, Dubna, Russia\\
$^{kb}$ Skobeltsyn Institute of Nuclear Physics, Lomonosov Moscow State University, Russia\\
$^{kc}$ Space Regatta Consortium, Korolev, Russia\\
% Slovakia ready (30032012)  - 2 institutes 
$^{la}$ Institute of Experimental Physics, Kosice, Slovakia\\
$^{lb}$ Technical University Kosice (TUKE), Kosice, Slovakia\\
% Spain ready (02042012)  - 5 institutes 
$^{ma}$ Universidad de Le\'on (ULE), Le\'on, Spain\\
$^{mb}$ Instituto Nacional de T\'ecnica Aeroespacial (INTA), Madrid, Spain\\
$^{mc}$ Universidad de Alcal\'a (UAH), Madrid, Spain\\
$^{md}$ Universidad Polit\'ecnia de madrid (UPM), Madrid, Spain\\
$^{me}$ Instituto de Astrof\'isica de Canarias (IAC), Tenerife, Spain\\
% Sweden ready (December 2013)  - 1 institutes 
$^{na}$ KTH Royal Institute of Technology, Stockholm, Sweden\\
% Switzerland ready (02042012) - 3 institutes 
$^{oa}$ Swiss Center for Electronics and Microtechnology (CSEM), Neuch\^atel, Switzerland\\
$^{ob}$ ISDC Data Centre for Astrophysics, Versoix, Switzerland\\
$^{oc}$ Institute for Atmospheric and Climate Science, ETH Z\"urich, Switzerland\\
% USA ready (30032012) - 9 institutes 
$^{pa}$ Space Science Laboratory, University of California, Berkeley, CA, USA\\
$^{pb}$ University of Chicago, IL, USA\\
$^{pc}$ Colorado School of Mines, Golden, CO, USA\\
$^{pd}$ University of Alabama in Huntsville, Huntsville, AL; USA\\
$^{pe}$ Lehman College, City University of New York (CUNY), NY, USA\\
$^{pf}$ NASA Marshall Space Flight Center, Huntsville, AL, USA\\
$^{pg}$ University of Utah, Salt Lake City, UT, USA\\
$^{ph}$ Georgia Institute of Technology, USA\\
$^{pi}$ University of Iowa, Iowa City, IA, USA\\
$^{pj}$ NASA Goddard Space Flight Center, Greenbelt, MD, USA\\
$^{pk}$ Center for Space Science \& Technology, University of Maryland, Baltimore County, Baltimore, MD, USA\\
$^{pl}$ Department of Astronomy, University of Maryland, College Park, MD, USA\\
$^{pm}$ Fairfield University, Fairfield, CT, USA
}

\vspace*{0.5cm}
\vspace*{0.5cm}

\newpage
{\small
\section*{Standard JEM-EUSO Acknowledgment in full-author papers}

This work was partially supported by Basic Science Interdisciplinary Research Projects of RIKEN and JSPS KAKENHI Grant (22340063, 23340081, and 24244042), by the Italian Ministry of Foreign Affairs and International Cooperation, by the Italian Space Agency through the ASI INFN agreements n. 2017-8-H.0 and n. 2021-8-HH.0, by NASA awards 11-APRA-0058, 16-APROBES16-0023, 17-APRA17-0066, NNX17AJ82G, NNX13AH54G, 80NSSC18K0246, 80NSSC18K0473, 80NSSC19K0626, 80NSSC19K0627 and 80NSSC18K0464 in the USA, by the French space agency CNES, by the Deutsches Zentrum f\"ur Luft- und Raumfahrt, the Helmholtz Alliance for Astroparticle Physics funded by the Initiative and Networking Fund of the Helmholtz Association (Germany), by Slovak Academy of Sciences MVTS JEM-EUSO, by National Science Centre in Poland grants 2017/27/B/ST9/02162 and 2020/37/B/ST9/01821, by Deutsche Forschungsgemeinschaft (DFG, German Research Foundation) under Germany's Excellence Strategy - EXC-2094-390783311, by Mexican funding agencies PAPIIT-UNAM, CONACyT and the Mexican Space Agency (AEM), as well as VEGA grant agency project 2/0132/17, and by by State Space Corporation ROSCOSMOS and the Interdisciplinary Scientific and Educational School of Moscow University "Fundamental and Applied Space Research".
}

\end{document}